\documentclass[article,structabstract]{aa}
\usepackage{graphicx}
\usepackage{amsmath}
\usepackage{txfonts}
\usepackage{xspace}
\usepackage{color}
\usepackage{natbib}
\bibpunct{(}{)}{;}{a}{}{,} % to follow the A&A style
\newcommand{\um}{\ensuremath{\rm \mu m}\xspace}
\newcommand{\e}[1]{\ensuremath{\rm ~\times~10^{#1}}\xspace}
\newcommand{\ten}[1]{\ensuremath{\rm 10^{#1}}\xspace}
\begin{document}

\title{Planet or Brown Dwarf? Inferring the Companion Mass in HD 100546 from the Wall Shape using Mid-Infrared Interferometry.}
\titlerunning{Planet or Brown Dwarf?}
 \author{Gijs D.~Mulders\inst{1,2}
\and Sijme-Jan Paardekooper\inst{3}
\and Olja Pani\'{c}\inst{4}
\and Carsten~Dominik\inst{1,5}
\and Roy van Boekel\inst{6}
\and Thorsten Ratzka\inst{7}
}

\institute{
%1
Astronomical Institute ``Anton Pannekoek'', University of Amsterdam,
 PO Box 94249, 1090 GE Amsterdam, The Netherlands
\and %2
SRON Netherlands Institute for Space Research, PO Box 800, 9700 AV,
Groningen, The Netherlands
\and %3
DAMTP, University of Cambridge, Wilberforce Road, Cambridge CB3 0WA, United Kingdom
\and %4
Institute of Astronomy, University of Cambridge, Madingley Road, Cambridge CB3 0HA, United Kingdom
\and %5
Department of Astrophysics/IMAPP, Radboud University Nijmegen, P.O. Box 9010, 6500 GL Nijmegen, The Netherlands
\and %6
Max-Planck Institute for Astronomy, K\"{o}nigstuhl 17, 69117 Heidelberg, Germany
\and %7
Universit\"{a}ts-Sternwarte M\"{u}nchen, Ludwig-Maximilians-Universit\"{a}t, Scheinerstr. 1, 81679 M\"{u}nchen, Germany
} % end institutes

%\date{ }
\offprints{G.D.Mulders, \email{\bf{mulders@uva.nl}}}

\abstract{%context
  Giant planets form in protoplanetary disks while these disks are still gas-rich, and can reveal their presence through the annular gaps they carve out. HD 100546 is a gas-rich disk with a wide gap between between a radius of $\sim$1 and 13 AU, possibly cleared out by a planetary companion or planetary system.
 }{%aims
   We want to identify the nature of the unseen companion near the far end of the disk gap.
}{%methods
  We use mid-infrared interferometry at multiple baselines to constrain the curvature of the disk wall at the far end of the gap. We use 2D hydrodynamical simulations of embedded planets and brown dwarfs to estimate viscosity of the disk and the mass of a companion close to the disk wall. 
}{%results
  We find that the disk wall at the far end of the gap is not vertical, but rounded-off by a gradient in the surface density. Such a gradient can be  reproduced in hydrodynamical simulations with a single, heavy companion ($\gtrsim$30...80 $M_{\rm Jup}$) while the disk has viscosity of at least $\alpha \gtrsim 5 \cdot 10^{-3}$. Taking into account the changes in the temperature structure after gap opening reduces the lower limit on the planet mass and disk viscosity to 20 $M_{\rm Jup}$ and $\alpha = 2 \cdot 10^{-3}$).
}{%conlusions
 The object in the disk gap of HD~100546 that shapes the disk wall is most likely a 60$^{+20}_{-40} M_{\rm Jup}$ brown dwarf, while the disk viscosity is estimated to be at least $\alpha = 2 \cdot 10^{-3}$. The disk viscosity is an important factor in estimating planetary masses from disk morphologies: more viscous disks need heavier planets to open an equally deep gap.
}

\maketitle

\section{Introduction}\label{sec:Introduction}
Giant planets need to form before the gas in protoplanetary disks is dispersed, thus making some of these disks not only planet-forming, but most likely also planet-hosting. Current planet-finding techniques have difficulties in detecting planets around these young stars: transits are blocked from view by the disk, whereas radial velocity measurements are disturbed by variability of the stellar photosphere \citep{2008Natur.451...38S,2008A&A...489L...9H}, though (interferometric) imaging has identified a few possible companions \citep{2012ApJ...745....5K,2013ApJ...766L...1Q}.

However, planets can also reveal themselves in an indirect way, through their dynamical impact on the protoplanetary disk \cite[e.g.][]{1979MNRAS.186..799L,1986ApJ...307..395L}. A gap carved by a single planet has a minimal impact on the SED, but could be identified by imaging of the disk \citep{2003ApJ...583L..35S,2006ApJ...637L.125V,2007P&SS...55..569W}. Nonetheless, a class of so-called transitional disks have been identified on the basis of their low near-infrared excess, apparently caused by depleted inner regions \citep{1989AJ.....97.1451S,2004ApJ...617..406M,2005ApJ...630L.185C}. Long-wavelength imaging has confirmed that most of these indeed have enlarged inner holes or annular gaps \citep{2009ApJ...704..496B,2011ApJ...732...42A}, their size suggestive that multiple planets must be responsible \citep{2011ApJ...738..131D,2011ApJ...729...47Z} or an additional clearing mechanism is at work, such as dust filtration or grain growth \citep{2012ApJ...755....6Z,2012A&A...544A..79B}. However, how these transitional disks manage to sustain a substantial accretion rate with a depleted inner region remains a mystery, as well as the underlying architecture of their planetary systems.

If present, the properties of an underlying planetary system may be inferred from the disk geometry. Giant planets will carve out deep and wide gaps in the gas and dust, and a whole suite of codes exists to study disk-planet interactions (see \citealt{2006MNRAS.370..529D} for a comparison of such codes). Comparing these results to observational constraints on the surface density profile allows to estimate the mass and location of a planet \citep[e.g.][]{2011A&A...531A...1T}.

In this work, we will study the disk-planet interaction in the disk of HD 100546. It was identified by \cite{2003A&A...401..577B} as a transitional disk on basis of its SED even before the term existed, with an estimated companion mass of $\sim 10 M_{\rm Jup}$. The gap was later confirmed using mid-infrared nulling interferometry \citep{2003ApJ...598L.111L}, UV spectroscopy \citep{2005ApJ...620..470G}, near-infrared CO spectroscopy \citep{2009ApJ...702...85B,2009A&A...500.1137V} and mid-infrared interferometry \citep{Panic:2012un}, all\footnote{At time of writing, there are no published observations that constrain the gap size at (sub)millimeter wavelengths.} consistent with a gap outer radius in the range of 10 to 15 AU. The inner disk is depleted in dust by a few orders of magnitude \citep{2010A&A...511A..75B,2011A&A...531A..93M}, and is less than 0.7 AU in size \citep{Panic:2012un}. Hydrodynamical modelling of the surface density profile inferred from the SED by \cite{2011A&A...531A...1T} yields a planet of at least one Jupiter mass at 8 AU. Recent interferometric imaging has revealed a (different) planetary candidate further out in the disk, at 70 AU \citep{2013ApJ...766L...1Q}.

However, the surface density profile around the gap can also be studied using mid-infrared interferometric data, providing additional constraints on the nature of a (planetary) companion. Using inclined ring models, \cite{Panic:2012un} have shown that the mid-infrared visibilities could not be reproduced by a sudden jump in intensity at the location of the disk wall at 13 AU, but that the emission from the edge increases smoothly over a few AU. This may indicate that the surface density does not show a sharp increase at 13 AU leading to a vertical wall, but gradually increases with radius. In this case the optical depth - which determines the height of the disk surface - also increases smoothly with radius, producing a more rounded-off wall. In this paper, we will explore the observational appearance of such surface density profiles, and how they can be explained by the gravitational and hydrodynamic interaction between a disk and a planet.

Such rounded-off walls are a natural outcome of hydrodynamical models of disk-planet interaction, where the detailed shape of the radial surface density profile depends on planet mass, disk thickness and viscosity \citep{2006Icar..181..587C,2006ApJ...641..526L}. The mass of the planet will therefore be reflected in the shape of the disk wall, that can be constrained with MIDI, the mid-infrared interferometer on the Very Large Telescope. We will explain this in more detail in Section \ref{sec:interf}. Besides the planets mass, the viscosity in the disk wall is a critical parameter in determining the gap shape.

We will use a 2D radiative transfer code to model both SED and mid-infrared visibilities to determine the shape of the surface density profile in the disk wall at the far end of the gap (Section \ref{sec:derive}). We then try to reproduce this surface density profile with a hydrodynamical code to constrain the planet mass and disk viscosity (Section \ref{sec:model}). We will discuss the companion mass and robustness of our result in Section \ref{sec:discussion} and summarize our results in the conclusion.

\section{Interferometric signature of a disk wall}\label{sec:interf}

\begin{figure}
  \includegraphics[width=\linewidth]{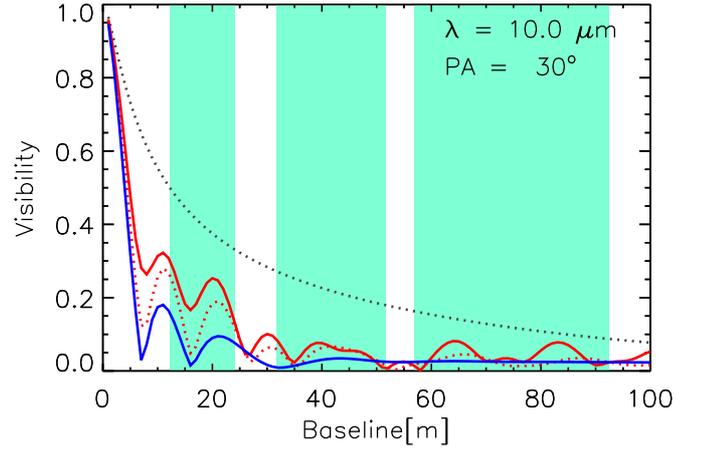}
  \caption[]{Visibilities versus baseline length for HD100546 at 10 micron, along a position angle of 30 degree (the same as the 41 meter baselines). The colored lines shown: a very round wall in blue, a steeper wall in dotted red, a vertical wall in solid red, and a model without a gap in dotted gray. Note that in the model without a gap the emission is coming from much closer to the star, and visibilities are much higher.
Indicated in green are the spatial frequency ranges probed by the MIDI spectra used in this paper ($\lambda=8...13$ micron), displayed as effective baseline at ten micron (${\rm Baseline}/\lambda \cdot 10\um$).
    \label{fig:bvis}}
\end{figure}

The visibility versus baseline curve is a Fourier transform of the surface brightness profile. Therefore, the visibility curve of a continuous disk\footnote{Assuming the inner and outer radius of the disk lie outside the emitting region, so there is no discontinuity in surface brightness at \textit{any} radius.} decreases monotonously with baseline (gray dotted line in figure \ref{fig:bvis}), as contributions from different radii add up to a smooth curve. In this case, the visibility is a direct measure of the spatial extent of a source.

\begin{figure}
  \includegraphics[width=\linewidth]{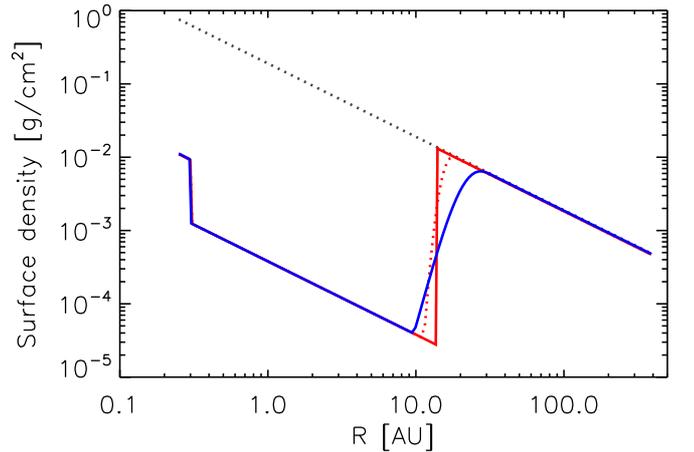}
  \caption[]{Surface density profile of the disk for the same models as in figure \ref{fig:bvis}. The surface density normalisations are those of the best-fit model described in table \ref{tab:model}.
    \label{fig:sdp}
}
\end{figure}

\begin{figure}
  \includegraphics[width=\linewidth]{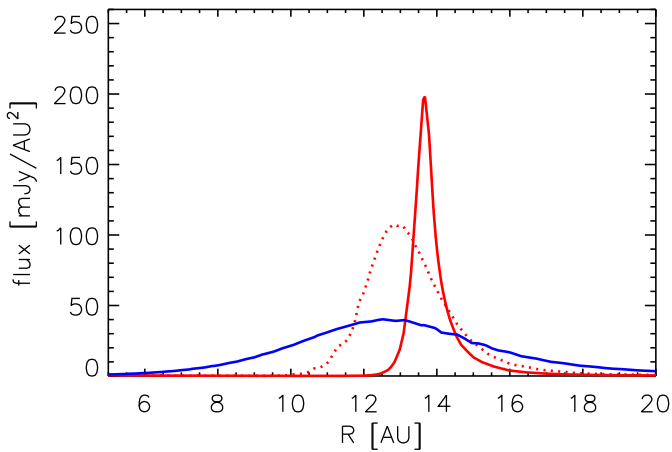}
  \caption[]{Radial brightness profile of the disk at 12.5 $\mu$m for the same models as in figure \ref{fig:bvis}. The model without a gap has no brightness peak in this range and is not shown.
    \label{fig:sbp}}
\end{figure}

However, in a disk with a gap, the situation is much more complex. The material at the far end of the gap (`disk wall') intercepts a much larger fraction of the stellar light than in a continuous disk, creating a peak in the surface brightness at that location, (red line in figure \ref{fig:sdp} and \ref{fig:sbp}). The flux contribution from this radius will then dominate over that of the other radii, both in the SED and visibilities. Its Bessel function will contribute more to the visibilities than those from other radii, resulting in the characteristic gap signature in the visibility curve, showing multiple minima (`bounces' or `nulls') and maxima (`sidelobes') (Figure \ref{fig:bvis}, red line).

The detailed shape of the visibility curve of a gapped disk depends on the structure of the disk wall: A vertical wall (defined as a step function in the surface density) will create a narrow peak in the surface brightness profile with a typical width of a few AU. This non-zero width is due to inclination and optical depth effects caused by the vertical structure. It has much more power in the sidelobes, making the gap structure visible at very long baselines (red lines in figure \ref{fig:bvis},\ref{fig:sdp} and \ref{fig:sbp}).
If the surface density increases gradually over a few AU (Fig. \ref{fig:sdp}, blue line), the optical depth will increase more slowly, giving a surface height in the disk wall that slowly increases with radius, rounding off the wall. Such a round wall will create a broader peak in the surface brightness profile (Fig. \ref{fig:sbp}, blue line). As shown in \cite{Panic:2012un}, this puts less power in the sidelobes, as overlapping Bessel functions cancel each other out, resulting in a smoother curve where the gap signature is not visible at longer baselines (Fig. \ref{fig:bvis}, blue line).

The shape of the disk wall at the far end of the gap can therefore be directly derived from the amount of structure in the visibility curve at long baselines. This method has been applied in the near-infrared to study the shape of inner rims at the dust sublimation radius \citep[e.g.][ their figure 5]{2008ApJ...689..513T,2010ARA&A..48..205D}. Indications for a round wall are also found with mid-infrared interferometry in the disk of \object{TW Hya} \citep{2007A&A...471..173R}.

In this work, we will use this method to derive the shape of the surface density profile in the disk wall from the MIDI observations presented in \cite{Panic:2012un} and \cite{2004A&A...423..537L}. The surface density profile directly affects the surface brightness profile and visibility curves (figures \ref{fig:bvis},\ref{fig:sdp} and \ref{fig:sbp}). Even though the MIDI observations used in this paper have a limited sampling of possible baselines, they still cover a considerable range in spatial frequency ($B/\lambda$) -- shown by the green areas in figure \ref{fig:bvis} -- and we can still perform such an analysis.

\begin{figure}
  \includegraphics[width=\linewidth]{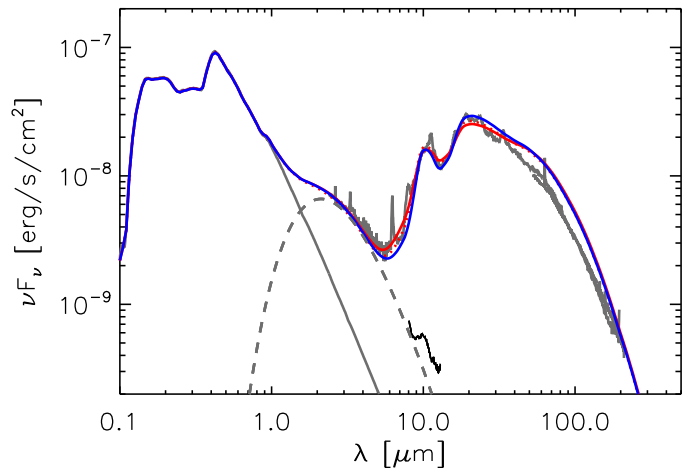}
  \caption[]{Spectral Energy Distribution of HD 100546. Plotted in gray are the ISO spectrum \citep{1998A&A...332L..25M} and the stellar photosphere \citep{1998A&A...330..145V}. Plotted in black is the 'inner disk spectrum' (the correlated spectrum of the 41.4 m baseline), as well as a 1700 K black body scaled to the K band flux (dashed line).
The colored lines show the same models as figure \ref{fig:bvis}. The model without a gap is not shown.
    \label{fig:sed}}
\end{figure}

\section{Deriving the wall shape}\label{sec:derive}
In this section we will constrain the shape of the surface density profile in the disk wall from the visibilities presented by \cite{Panic:2012un} and \cite{2004A&A...423..537L}. To do so, we need to first calculate the density and temperature structure in the disk wall for a given surface density profile. Because the structure of a rounded disk wall can strongly deviate from that of a geometrically thin disk, this is a job especially well suited for 2D radiative transfer codes - also because the disk wall near the midplane is completely shadowed by the inner disk, while the upper part is fully illuminated. In addition, we need to calculate the vertical structure in the disk wall, which strongly deviates from that of a continuous disk due to radial and vertical temperature gradients.

We use {\tt MCMax} \citep{2009A&A...497..155M}, a 2D radiative transfer code that self-consistently calculates the temperature \textit{and} vertical density structure for a given surface density profile in an axisymmetric geometry. We assume the gas is in vertical hydrostatic equilibrium, and because the SED shows no sign of dust settling of small grains \citep{2013A&A...549A.112M}, we assume the dust and gas distributions to be equal. The SED and visibilities are calculated using ray-tracing to compare them to the observations. We will use the SED fit with a vertical wall presented in \cite{2011A&A...531A..93M} as a starting point, refine it using the constraints from \cite{Panic:2012un}, and fit the SED and visibilities simultaneously to constrain the surface density in the disk wall. For completeness, all model parameters are displayed in table \ref{tab:model}.

      \begin{table}
        \centering
        \title{Model parameters}
        \begin{tabular}{lll}
          \hline \hline
          Parameter  & Value & reference \\
          \hline  %% star
          $T_{\rm eff} ~[{\rm K}]$      & 11000  & [1] \\
          $L_* ~[L_{\odot}]$  & 33  & [1] \\
          $M_* ~[M_{\odot}]$  & 2.4   & [1] \\
          $d ~[{\rm pc}]$ & 97 & [2] \\
          \hline  %% radii
          $r_{\rm in,inner} ~[{\rm AU}]$  & 0.25  & [3] \\
          $r_{\rm out,inner} ~[{\rm AU}]$ & 0.3 & $\dagger$ \\
          $r_{\rm out,outer} ~[{\rm AU}]$ & 400 & [4] \\
          $r_{\rm exp} ~[{\rm AU}]$ & 29 & $\dagger$ \\
          \hline %% mass
          $M_{\rm dust} [M_{\odot}]$ & 5\e{-5} & $\dagger$ \\
          Vertical structure & hydrostatic & \\
          \hline %% gap structure
          $\Sigma_{\rm inner}(r) ~[{\rm g/cm}^2]$  & $0.003 ~(r/{\rm AU})^{-1}$ & $\dagger$  \\
          $\Sigma_{\rm gap}(r) ~[{\rm g/cm}^2]$  & $0.0004 ~(r/{\rm AU})^{-1}$& $\dagger$   \\
          $\Sigma_{\rm wall}(r) ~[{\rm g/cm}^2]$ & see Eq. \ref{eq:shape}& $\dagger$   \\
          $\Sigma_{\rm outer}(r) ~[{\rm g/cm}^2]$ & $0.2 ~(r/{\rm AU})^{-1}$ & $\dagger$   \\
          \hline %%dust
          $a_{\rm min}[\um]$ & 0.1 & [5] \\
          $a_{\rm max}[\um]$ & 1.5 & [5] \\
          Shape  & irregular (DHS) & [5] \\
          Silicate fraction $[\%]$ & 70  & $\dagger$ \\
          Carbon fraction $[\%]$ & 30 & $\dagger$ \\
          \hline %% i, PA
          $i [\degr]$ & 53 & [6] \\
          PA & 145 & [6] \\
          \hline \hline
        \end{tabular}
        \caption{Parameters with a dagger (${\dagger}$) are (re)fitted.
          References:
          [1] \cite{1997A&A...324L..33V};
          [2] \cite{2007ASSL..350.....V};
          [3] \cite{2010A&A...511A..75B};
          [4] \cite{2007ApJ...665..512A};
          [5] \cite{2010ApJ...721..431J};
          [6] \cite{Panic:2012un}; 
          Silicate composition:     
          10\% MgFeSiO$_{4}$, 
          28\% MgSiO$_{3}$, 
          31\% Mg$_{2}$SiO$_{4}$, 
          1\%  NaAlSi$_{2}$O$_{6}$ (Optical constants from \citealt{1995A&A...300..503D,1996A&A...311..291H,1998A&A...333..188M}.
          Optical constants for Carbon from \citep{1993A&A...279..577P}.
          \label{tab:model}}
      \end{table}

\begin{figure}
  \includegraphics[width=\linewidth]{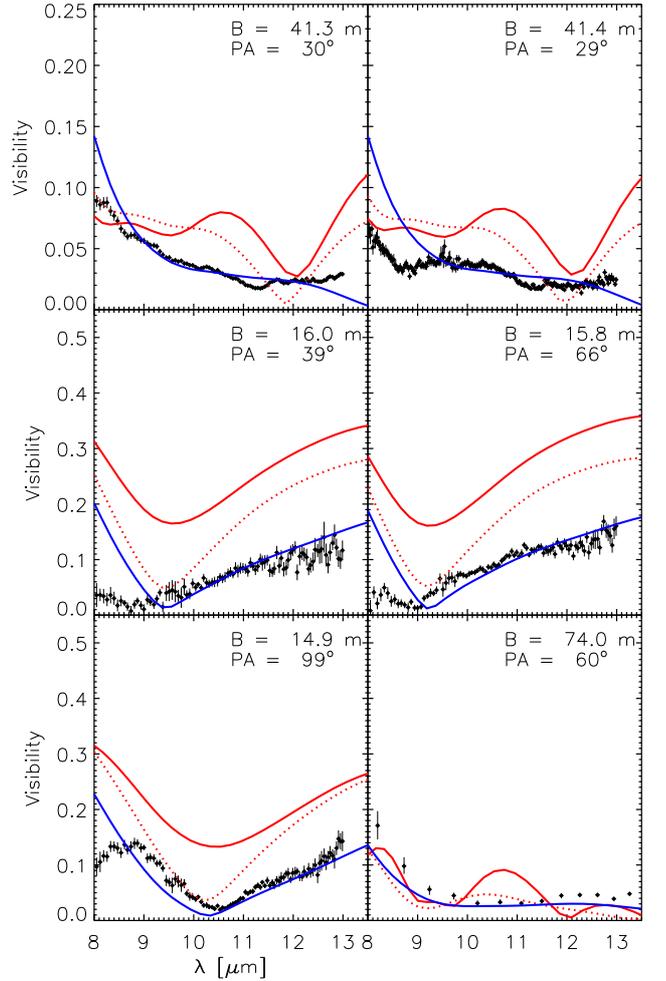}
  \caption[]{Observed visibilities for HD100546 on different projected baseline lengths and orientations (diamonds). We use the total flux from ISO to calculate visibilities from the correlated fluxes. The colored lines show the same models as figure \ref{fig:bvis}. The model without a gap is not shown.  
    \label{fig:vis}}
\end{figure}

To generate visibilities from the observed correlated fluxes, we need to divide them by an observed total flux. Because the total flux measured with MIDI is affected by flux losses from the MIDI slit (see \citealt{Panic:2012un}), we use the flux measured with ISO to generate visibilities. The larger field of view of ISO also includes more large-scale PAH emission, with narrow features around 7.9, 8.6 and 11.3 \um, which are inversely imprinted in the visibilities, but which we do not model. These visibilities, together with those presented by \cite{2004A&A...423..537L}, are shown in Figure \ref{fig:vis}. We will use the position angle of 145\degr{} and inclination of 53\degr{} derived by \cite{Panic:2012un} to compute visibilities from our disk models.

\subsection{Inner disk}\label{sec:inner}
As shown by \cite{Panic:2012un}, the inner disk is -- at least in the small dust grains probed by MIDI -- very compact, $<0.7$ AU . This is perhaps best illustrated by plotting the correlated flux on the 41 meter baseline into the SED of figure \ref{fig:sed}. This baseline probes scales of order $\sim$2...3 AU, and therefore filters out most of the emission from the outer disk. The correlated flux seems consistent with the 1700 K blackbody emission from the inner rim, with only little additional flux from colder material. We place the inner rim at 0.25 AU, consistent with the near-infrared interferometric constrains from \cite{2010A&A...511A..75B} and \cite{2011A&A...531A...1T}. Because the temperature behind the inner rim at 0.25 AU drops off very rapidly with radius, we find that the inner disk can not extend further out than 0.3 AU (in small grains), unless the surface density power law is steeper than $r^{-1}$.  This tiny inner disk is similar to that in the transitional disk \object{T Cha} \citep{2013A&A...552A...4O}. The normalization of the surface density ($\Sigma_{\rm inner}$) is fitted to the near-infrared excess.

In addition, there is a weak feature apparent at 10 micron, though it is not clear whether this is a mineralogical feature in the inner disk, or due to the structure of the outer disk. The outer disk wall also contributes to the correlated flux and creates similar features -- which we will discuss extensively in the next section. If we treat the observed feature strength as an upper limit to the real feature strength, the feature is too weak to be consistent with a rim made out of the same small amorphous silicate grains as the outer disk wall. Because the temperature of the inner rim is above the crystallisation temperature of silicates -- and possibly also above its sublimation temperature -- we use a composition of pure iron, which also fits the SED (Figure \ref{fig:sed}). We can mix in a few percent of silicates or corundum (FeAl$_2$O$_3$, which has a higher sublimation temperature) to fit the weak 10 micron feature in the correlated spectrum, but this is not  required for a good fit to the visibilities and has no effect on the results presented in this paper.

\subsection{A vertical wall}
As explained in section \ref{sec:interf}, the presence of a disk wall creates structures in the visibility as function of baseline, which will be reflected in the spectrally resolved visibilities as well (Fig. \ref{fig:vis}). This is most clearly seen in the 14.9 m baseline, where a bounce is present at 10.5 \um. Bounces are also present around 8.5 \um in the 15.8 and 16.0 m baselines. The longer baselines do not show such pronounced structures.

 A disk model with a vertical wall at 14 AU can reproduce the location of these bounces at the shortest baselines (Fig. \ref{fig:vis}, red lines). However, this model overestimates the visibilities at these baselines. In addition, the vertical wall model predicts structures at the 41 and 74 meter baselines that are not observed. To fit the visibilities at all baselines, we have to round off the disk wall as described in section \ref{sec:interf}. This reduces the power in the sidelobes, thus reducing the visibilities at the shortest baselines and smoothing out structures at the longer baselines.

\subsection{A rounded-off wall}
We round off the disk wall by modifying its surface density power law, using the following function:
\begin{equation}\begin{split}\label{eq:shape}
  \Sigma_{\rm wall}(r <   r_{\rm exp}) ~=~& \Sigma_{\rm outer}~r^{-1}~\exp\left( -\left(\frac{1-r/r_{\rm exp}}{w}\right)^3\right)\\
  \Sigma_{\rm disk}(r \geq r_{\rm exp}) ~=~& \Sigma_{\rm outer}~r^{-1}
\end{split}\end{equation}
where $\Sigma_{\rm outer}$ is  the normalisation constant for the outer disk surface density fitted to the SED, $r_{\rm exp}$ is the radius where the drop-off of the surface density sets in, and $w$ is a measure of how round the disk wall is. This function is essentially a Gaussian similar to the vertical scale height, but with a different exponent of 3 as in \cite{2006ApJ...641..526L}, eq. 5. Both $w$ and $r_{\rm exp}$ are free fitting parameters. Examples of the resulting surface density distributions are shown in figure \ref{fig:sdp}. Although different parametrizations than the one introduced here are possible, this particular one is chosen because it also accurately describes the outcome of our hydrodynamical simulations in the next section. 

We first round off the disk wall using $w=0.20$ and $r_{\rm exp}=19$ AU (Fig. \ref{fig:sdp}, red dotted line), consistent with the shape of the disk wall as modelled by \cite{2011A&A...531A...1T}. Although this produces a smoother surface brightness profile than a vertical wall (Fig. \ref{fig:sbp}) with less power in its sidelobes (Fig. \ref{fig:bvis}), it still produces too much structure on the 41 and 74 meter baselines and overpredicts the shortest baselines. To completely remove all structure on the longer baselines, we have to round off the disk wall even further. Our best fit model has $w=0.36$ and $r_{\rm exp}=29$ AU, shown by the blue lines in figures \ref{fig:bvis}, \ref{fig:sdp}, \ref{fig:sbp}, and \ref{fig:vis}. There is a small range of solutions with almost equally good fits, ranging from $w=0.33$ and $r_{\rm exp}=26$ AU to $w=0.40$ and $r_{\rm exp}=35$ AU.  The total parameter space explored ranges from w=0.00 to w=0.60, and from $r_{\rm exp}=10$ to $40$ AU.

We note that due to disk asymmetries and variability as discussed by \cite{Panic:2012un}, it is impossible to find a perfect fit to all data with a single axisymmetric model. Especially the 8...9 \um region dominated by the inner disk is affected by this, as observations taken at the same projected baseline length and position angle but at a different date have a different shape. However, the structures arising from vertical walls are not seen in any of the observations, so we are confident that our results for the wall shape are robust. Fitting each baseline separately does provide better fits, but does not change our fit parameters by more than $\Delta w=0.02$ and $\Delta r_{\rm exp}=2$ AU.

\section{Deriving planet mass and disk viscosity}\label{sec:model}
In this section, we will show how a planet can explain the observed shape of the disk wall. The micron sized-dust grains observed with MIDI are a good tracer of the gas: they are well-coupled to the gas \citep[][their section 4.2]{2013A&A...549A.112M}, unlike millimeter sized grains that tend to pile up near pressure bumps in the midplane \citep{2004A&A...425L...9P,2012A&A...545A..81P}. Therefore, the inferred surface density of the dust is equal to that of the gas, which we will reproduce using planet-disk interactions.

\subsection{Gap opening}\label{sec:open}
A planet embedded in a disk opens a gap by exerting a torque on the disk, pushing material outside of its orbit outwards and material inside of it inwards. However, gas flows back into the gap due to pressure gradients and viscous spreading, which tends to close the gap. Therefore, the shape of the surface density around a planet depends on its mass, the disk viscosity and scale height \citep{2006Icar..181..587C,2006ApJ...641..526L}. In general, a heavier planet can carve out a deeper and wider gap, while a more viscous or thicker disk (higher pressure scale height) will reduce the gap width and depth. 

\subsection{Hydrodynamical model}
We use the freely-available 2D hydrodynamical code {\tt Fargo} \citep{2000ASPC..219...75M} to model the surface density density of the  gas disk around an embedded  planet's orbit. {\tt Fargo} is a solver to the Navier-Stokes and continuity equations in a differentially rotating disk in vertical hydrostatic equilibrium. It speeds up these calculations by removing the average azimuthal velocity component from these equations at each radius and at each time-step.

Because we focus on the steady-state wall shape, it is not necessary to follow the global evolution of the whole disk as \cite{2011A&A...531A...1T} have done. Instead, we focus on the surface density profile in the vicinity of the planet. We model the disk within a factor 5 in radius of the planet (i.e., 2 to 50 AU for a planet at 10 AU)  far enough from the grid edges so that their location do not affect the shape of the surface density in the vicinity of the planet. The planetary candidate discovered by \cite{2013ApJ...766L...1Q} lies outside this grid, and -- assuming it is on a circular orbit -- is therefore located too far away to influence the gap structure. We use square grid cells on a logarithmic grid , 384 in the azimuthal direction. We use a non-reflective boundary condition to suppress wave-reflection at the inner boundary and prevent mass from leaking out at the inner edge of the grid. 

The model set-up is based on our best-fit radiative transfer model to the SED and visibilities. We use a surface density power law of $r^{-1}$ for the initial surface density, similar to that of the outer disk. We use a pressure scale height derived from the scale height profile of our radiative transfer simulation from \textit{within} the disk gap , i.e. between 0.3 and $\sim$13 AU. Because our radiative transfer model is not vertically isothermal, the scale height is defined as the height above the midplane where the pressure drops off by $e^{-1/2}$. The pressure scale height in this regime is well-fitted by a power law of the form $H_{\rm p}(r)= 0.025~r^{1.39}$, see also Figure \ref{fig:Tprof}. We will discuss the influence of the chosen scale height profile in the discussion (Section \ref{sec:Tprof}).

To describe the viscous evolution of the disk, we use a viscosity of $\alpha$ type \citep{1973A&A....24..337S,1981ARA&A..19..137P},  consistent with our choice of $r^{-1}$ for the initial surface density profile. For the long integration times required to reach a steady state, such a viscosity agrees well with simulations of magnetohydrodynamical turbulence \citep{2004MNRAS.350..829P}. We explore the range from  $\alpha=10^{-4}$ to $\alpha=5\times 10^{-2}$, above which the time step calculation of {\tt Fargo} may no longer be correct.

\begin{figure}
  \includegraphics[width=\linewidth]{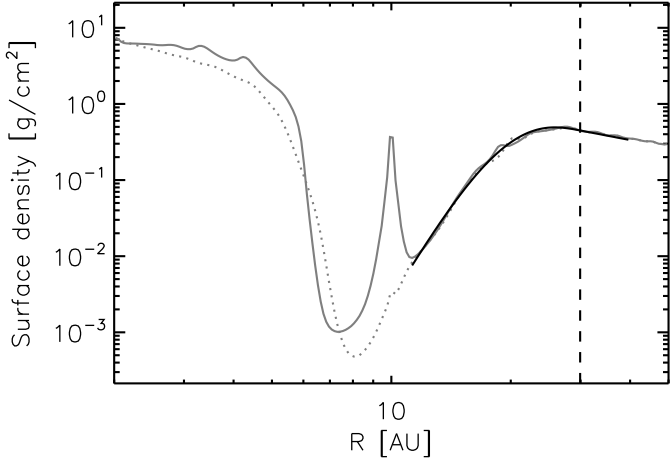}
  \caption[]{Surface density around a 60 Jupiter mass planet in a viscous ($\alpha=2\e{-2})$ disk after $10^4$ orbits, with and without accretion onto the planet (dotted  and solid gray line, respectively). The solid black line is the analytical fit using equation \ref{eq:shape} with parameters that also fit the MIDI data ($w=0.36$). The dashed line denotes $r_{\rm exp}$, the radius where the surface density starts to deviate from a power law.
    \label{fig:shape}}
\end{figure}

We consider a wide range of planets from 1 Jupiter mass to the hydrogen burning limit at 80 Jupiter masses. The reason for considering such high planet masses is that higher viscosities tend to close the gap, requiring much heavier planets to keep the gap open. We follow the disk evolution for $10^4$ orbits of the planet around a 2.4 solar mass star (2\e{5} yr at 10 AU), after which all models have reached a steady-state. The planet is assumed to be on a circular orbit, and to neither migrate nor accrete\footnote{With the exception of the model represented by the dotted line in figure \ref{fig:shape}, which uses the maximum accretion efficiency as defined in \cite{1999MNRAS.303..696K}.}

\subsection{Gap shape}\label{sec:shape}
To compare the surface density profiles produced by {\tt Fargo} with our radiative transfer models, we use the analytical fit function of equation \ref{eq:shape}. This function is fitted to the surface density profile  outside the radial location of the planet, between $r_{\rm min}$ and $r_{\rm max}$, where $r_{\rm min}$ is the location of the minimum gap depth outside the planets orbit ($r_{\rm min}>r_{\rm p}$) , and $r_{\rm max}$ is taken near the edge of the grid at $4.5 r_{\rm p}$. This radius is chosen to be just inside the outer grid edge, to avoid the region where the spiral wake hits this edge and local boundary effects may be important. 

This function fits the surface density profiles to an accuracy of about 10\%. An example is shown in figure \ref{fig:shape}. The main deviations come from transient features at small  spatial scales, while the overall shape is generally well produced.  Note that we do not use the results of the hydrodynamical simulation inside the planets orbit, and instead assume this region is empty as indicated by our radiative transfer model, see Section \ref{sec:width} for a discussion on the gap width.

In addition, we measure the depth of the gap by taking the minimum value of the density with respect to the unperturbed density ($\Sigma_{\rm outer}~r^{-1}$). Note that this is only a lower limit to the real depth of the gap: material that is corotating with the planet or orbiting around it creates a surface density spike at the location of the planet. If planetary accretion is turned on  , using the maximum accretion efficiency following the prescription of \cite{1999MNRAS.303..696K}, the spike disappears and the gap becomes deeper, typically by a factor of 2, though the shape of the disk wall at the far end of the gap does not change significantly, as shown by the dotted line in figure \ref{fig:shape}.

\begin{figure}
  \includegraphics[width=\linewidth]{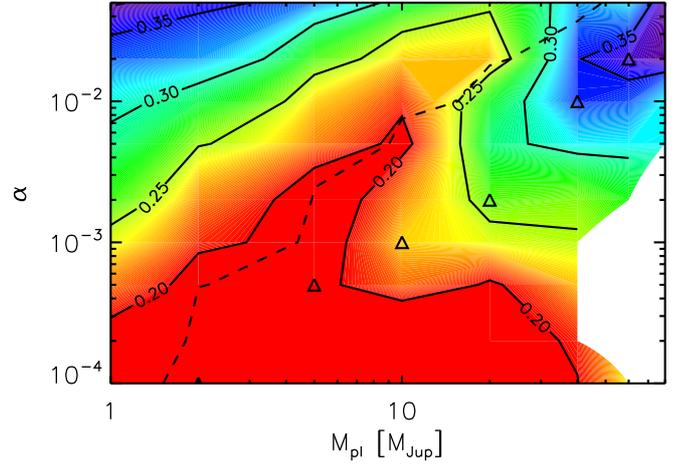}
  \caption[]{Shape of the disk wall as function of planet mass and disk viscosity. Solid contours and colors denote fitting parameter $w$ from equation \ref{eq:shape}, which is $w\sim$0.33...0.40 for our best-fit radiative transfer models. The dashed line denotes a gap depth of $\ten{-3}$. In the region below, the gap is deep enough to be consistent with the observed visibilities. Blue colors indicate rounder walls, red colors more vertical walls. The triangles indicate the models used for iterating on the pressure profile (See section \ref{sec:Tprof}).  The white regions contain models that neither converged nor finished $10^{5}$ orbits due to excessive computing time.
    \label{fig:contour}}
\end{figure}

\subsection{Results}
The results of our parameter study are presented in figure \ref{fig:contour}, showing the width over which the surface density in the disk wall is rounded off ($w$ from equation \ref{eq:shape}) as a function of planet mass and disk viscosity. A global trend is visible, where lower planet masses and disk viscosities produce more vertical walls ($w \sim 0.2$), whereas rounder walls are found at higher values ($w \sim 0.35$). This trend can be understood from a balance between viscous spreading and planetary torques: a high viscosity makes material flow inward smoothing the surface density profile, while a heavier planet mass allows the torques to act over a wider range, allowing for a shallower profile. However, for intermediate shapes there is no linear trend ($w \sim 0.25...0.3$).

There is an additional observational constraint on the surface density profile, namely the depth of the gap \citep{2011A&A...531A...1T}. If the gap is not deep enough, its emission will fill the gap in the SED around 8 micron and will over predict all visibilities. Because the inner disk is already depleted by a factor of a hundred, the gap needs to be depleted by about a factor of about ten more (See figure \ref{fig:sdp}), depending on the dust opacities. This minimum gap depth is shown with the dashed line in figure \ref{fig:contour}, and only models below this line have gaps that are deep enough. It shows that the disk viscosity is a crucial parameter in opening a disk gap: a Jupiter mass planet may open a deep gap in an inviscid disk, but for the most viscous disks it requires a 80 Jupiter mass planet to keep the  gap deep enough. In general, we find that the planet mass needs to be scaled with the square root of viscosity to achieve a given gap depth. 

There are two regions in this diagram which have a wall shape consistent with our observations, $w\sim0.33...0.40$. One region lies at low planetary masses ($<5 M_{\rm Jup}$) and high viscosities ($\alpha > 2\e{-2}$). Although consistent with the planet mass estimate of \cite{2011A&A...531A...1T}, the higher viscosity acts against gap opening, and these gaps are nowhere near deep enough to be consistent with observations, even if the planet would be allowed to accrete. Another region lies at very high masses ($30...80 ~M_{\rm Jup}$) and moderately high viscosities ($\alpha = 5\e{-3} ~... ~5\e{-2}$), which seems consistent with the observed gap depth, and puts the planet between 8 and 10 AU. Our best fit-model has a planet of $60~M_{\rm Jup}$ at 10 AU and a viscosity of $\alpha = 2\e{-2}$, and is shown in figure \ref{fig:shape}. The dependence of these results on the assumed temperature profile is discussed in Section \ref{sec:Tprof}.

These estimates of $\alpha$ are consistent with that of \cite{2013A&A...549A.112M}, who also find a high turbulent mixing strength of $\alpha_{\rm turb} > 0.01$ by looking at the degree of dust settling in the disk wall.

\section{Discussion}\label{sec:discussion}

\subsection{Planet or brown dwarf?}
The shape of the disk wall points to a companion about thirty Jupiter masses or heavier, with a best fit around 60 Jupiter masses. According to the IAU definition, this object would be a brown dwarf, rather than a planet, making HD100546 a misinterpreted binary system with a disk like CoKu Tau/4 \citep{2008ApJ...678L..59I}, rather than a transitional disk.

Binaries are common around main sequence A stars and Herbig Ae/Be stars, with fractions over 70\% \citep[e.g.][]{2006MNRAS.367..737B,2007A&A...474...77K}. The inferred period of around 20 years is consistent with the peak in the period distribution of binaries around sun like stars \citep[e.g.][]{1984Ap&SS..99...41Z,2010ApJS..190....1R}. Whether a mass ratio of $q \sim 0.02$ is uncommon for binaries is not clear, as detection limits typically go to $q \sim 0.1$, but \cite{2010MNRAS.401.1199W} note that the companion distribution of Herbig Ae/Be companions is skewed towards higher masses than the interstellar mass function. However, we note that the companion  candidate recently discovered around \object{HD 142527} by \cite{2012ApJ...753L..38B} has a very similar mass and orbital properties. In addition, there seems to be a trend from direct imaging that A type stars posses heavier planets than less massive stars \citep{2008Sci...322.1348M,2010Sci...329...57L,2013ApJ...763L..32C}, a trend also seen in radial velocity surveys \citep{2007A&A...472..657L}.

Whether to call a companion a planet or a brown dwarf should depend on how it forms, not on its mass. Young stars are found with masses below the deuterium burning limit \citep[e.g.][]{2005ApJ...635L..93L}, while core-accretion models predict that planets can also form above it \citep{2009A&A...501.1139M}. A better definition would involve the formation process of the companion, in a disk or like a single star. The current gas mass of the disk is uncertain due to lack of resolved millimeter observations, but estimated to be in the range $0.0005...0.01 M_\odot$ \citep{2010A&A...519A.110P,2011A&A...530L...2T} and therefore much lighter than the planet, making it unlikely that the planet recently formed out of the disk.

Observational limits exist on the mass and location of a possible binary companions of HD 100546. \cite{2006MNRAS.367..737B} report no detection in a spectroastrometric survey for binary companions, with a contrast limit of 6 magnitudes and and minimum separation of 0.1 \arcsec, excluding a companion with spectral type earlier than M6\footnote{Calculated at 5 Myr using the evolutionary tracks from \cite{2000A&A...358..593S}.}) outside of 10 AU. Limits also exist on a companion in the gap: \cite{2005ApJ...620..470G} used UV spectroscopy to put an upper limit to the spectral type later than M5, excluding a stellar companion, but not a brown dwarf. 

The projected location of the planet (0.05..0.1\arcsec{})  is just inside the region that can be surveyed with current direct imaging instruments \citep{2011ApJ...738...23Q}, and also falls within the region blocked by coronagraphs of upcoming planet-hunting instruments such as VLT/SPHERE and the Gemini Planet Imager. However, it is within the reach of techniques such as Sparse Aperture Masking \citep[e.g.][]{2012ApJ...745....5K,2012ApJ...753L..38B}. We have estimated the contrast ratio using the \cite{2002A&A...382..563B} evolutionary tracks. A 20 to 75 Jupiter mass brown dwarf with an age of less than 10 Myr has a luminosity between 0.03 and 0.3 $L_\odot$ and a temperature between 2500 and 3000 K. In the H, K and L band, this results in a contrast ratio between $5\e{-4}$ to $1\e{-2}$, or 5 to 8 magnitudes, within the detectable range.

\subsection{Gap Width}\label{sec:width}
The width of the gap from the hydrodynamical simulations, about 6 AU, is inconsistent with the derived size for the inner disk of less than an AU. The planet itself is heavier than the disk and should not migrate, allowing it to act as a barrier for material from the outer disk to reach the inner disk. We have neglected the global evolution of the disk, so it is possible that in reality, the inner disk drains onto the central star as in \cite{2011A&A...531A...1T}. The near-infrared interferometric data used by the authors, however, does not directly constrain the outer radius of the inner disk, which they place at $4$ AU. The MIDI data do require the inner disk to be much smaller than the $4$ AU previously assumed (see also \citealt{Panic:2012un}). Therefore, an additional mechanism might be necessary for clearing the inner regions. 

Additional planets closer to the star could explain the extent of the gap, depleting the inner regions \citep{2011ApJ...738..131D,2011ApJ...729...47Z}.  Because the torques on the disk are strongest close to a planet, we expect only the outermost planet in the gap to shape the disk wall. Another mechanism could be dust filtration at the outer edge, blocking dust particles from crossing the gap together with the gas and reducing the dust to gas ratio of the inner disk \citep{2012ApJ...755....6Z}. However, the high viscosities we infer are less favorable for trapping dust in the outer disk wall \citep{2012A&A...545A..81P}. In addition, grain growth could contribute to depleting the inner disk of small grains \cite{2012A&A...544A..79B}. Resolved millimeter observations of the gap, such as can be delivered with the Atacama Large Millimeter Array, would be crucial to investigate this.

\begin{figure}
  \includegraphics[width=\linewidth]{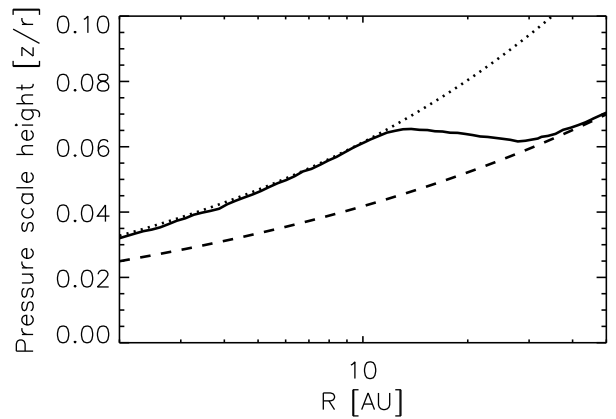}
  \caption[]{Scale height profile for our best fit radiative transfer model (solid line) in the range modelled with {\tt Fargo}. Also plotted are the best fit profile in the gap ($H_{\rm p}(r)= 0.025~r^{1.39}$, dotted line) and that of a disk without a gap ($H_{\rm p}(r)= 0.02~r^{1.32}$, dashed line). Because the radiative transfer model is not vertically isothermal, the scale height is defined as the height where the pressure drops off with a factor $e^{1/2}$ with respect to the midplane.
    \label{fig:sh}}
\end{figure}

\begin{figure}
  \includegraphics[width=\linewidth]{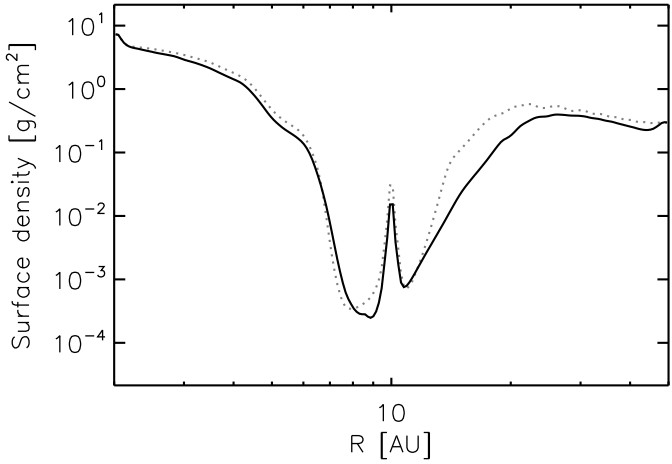}
  \caption[]{Surface density around a 20 Jupiter mass planet in a viscous ($\alpha=2\e{-3})$ disk for two different temperature profiles. The dotted line has a power-law scale height profile resulting in a steep wall ($w=0.26$), the solid line uses the scale height profile from the radiative transfer calculation, resulting in a rounder wall ($w=0.32$).
    \label{fig:Tprof}}
\end{figure}

\subsection{Pressure scale height}\label{sec:Tprof}
As mentioned in section \ref{sec:open}, the disk thickness -- set by the pressure scale height and thus by temperature -- is an important factor in determining the wall shape because it controls how much material flows back into the gap. The pressure scale height is an input parameter of {\tt Fargo}, and remains fixed during the simulation. This is a highly idealized situation, and in reality, the scale height will change during gap opening when the disk wall heats up \citep{2012ApJ...748...92T}. This will, in turn, influence the gap structure, but coupling the hydrodynamical simulations to radiative transfer is far from trivial. In this section we will describe how this coupling may affect our results and suggest directions for future work.

Because we are looking for a steady state solution of the wall shape after many orbits, we can derive the scale height profile corresponding to this wall shape from our best-fit radiative transfer model to the SED and visibilities. This profile is shown in figure \ref{fig:sh}. In the region up to $\sim$12 AU,  i.e. within the disk gap, the vertical optical depth is so low that the midplane is not shielded, increasing the temperature by a factor of two and the pressure scale height by about 50\% with respect to a disk without a gap (dashed line, see also \citealt{2012ApJ...748...92T}). Between $\sim$12 and 30 AU, the gradual increase in optical depth starts shielding the midplane, leading to lower temperatures and scale height. Outside 30 AU, the scale height profile is close to that of a continuous disk.

In the parameter study, we have used a parameterized scale height of the form $H_{\rm p}(r)= 0.025~r^{1.39}$ (Fig. \ref{fig:sh}, dotted line). This parametrization describes the profile accurately up to $\sim$12 AU, i.e. in most of the region modelled with {\tt Fargo}. In addition, in the region where the intensity profile peaks (see Fig. \ref{fig:sbp}) and where our observations are most sensitive to, it is closer to the scale height  of the best-fit radiative transfer model than that of a continuous disk is. However, the slope of the scale height profile after the peak in intensity at $\sim$12 AU is very different. In a steady-state viscous disk without a gap, it is this slope that sets the surface density distribution \citep[e.g.][]{2009ApJ...700.1502A}. To see how this assumption affects our results, we designed the following test.

\begin{table}
  \title{Wall shape and location}
  \centering
  \begin{tabular}{ll|ll|l|ll}
    \hline \hline
    \multicolumn{2}{c|}{model} & \multicolumn{2}{|c|}{before iteration} & \multicolumn{1}{|c|}{RT} & \multicolumn{2}{|c}{after iteration} \\
    
     $M_{\rm pl}$ & $\alpha$ & $w$  & $r_{\rm exp}$ [$r_{\rm pl}$] & $r_{\rm exp}$ [AU] & $r_{\rm pl}$ [AU] & $w$ \\
    \hline
    2  & $        10^{-4}$ & 0.17 & 1.6 & 17 & 10.6 & 0.21 \\
    5  & $5 \cdot 10^{-4}$ & 0.20 & 1.8 & 18 & 10.2 & 0.27 \\
    10 & $        10^{-3}$ & 0.23 & 2.0 & 19 &  9.6 & 0.28 \\
    20 & $2 \cdot 10^{-3}$ & 0.26 & 2.3 & 21 &  9.2 & 0.33 \\
    40 & $        10^{-2}$ & 0.35 & 2.9 & 28 &  9.8 & 0.36 \\
    60 & $2 \cdot 10^{-2}$ & 0.36 & 3.0 & 29 &  9.7 & 0.37 \\
    \hline \hline
  \end{tabular}
  \caption{Wall shape and location, before and after iterating on the temperature structure. The wall shape $w$ and its location $r_{\rm exp}$ are defined in equation \ref{eq:shape}. The location of the planet $r_{\rm pl}$ follows from comparing the fitted wall location in the radiative transfer code ($r_{\rm exp}$ in column RT) to that of the hydrodynamical simulation. This subset of models is highlighted by triangles in the parameter study of figure \ref{fig:contour}.
}
\label{tab:Tprof}
\end{table}

We perform one iteration on the hydrodynamical structure of the disk, to take into account the change in temperature structure of the disk after gap opening. We start with a subset of the hydrodynamical models presented in the previous section, shown in Table \ref{tab:Tprof} and indicated by triangles in figure \ref{fig:contour}. These models span the entire range of wall shapes, from steep to rounded-off, and were  initially calculated using the parameterized scale height profile. For each wall shape $w$, we calculate the temperature structure using our radiative transfer code, as described in section \ref{sec:derive}. The radial location of the wall depends -- apart from on the planets location -- also on the planet mass, because more massive planets carve out wider gaps. Therefore we adjust the location of the planet $r_{\rm pl}$ -- and hence that of the disk wall ($r_{\rm exp}$, Eq. \ref{eq:shape}) -- to make sure that the radial intensity profile peaks at 12-13 AU for each wall shape, as in figure \ref{fig:sbp} . This is equivalent to fitting the SED (but not the visibilities), since that only depends on the wall location, not its shape.
 
 From these temperature structures we can calculate the scale height profiles, similar to figure \ref{fig:sh}. We use these (non-parameterized) profiles to rerun the hydrodynamical simulations, and measure the  new wall shapes of these iterated models using equation \ref{eq:shape}. If the resulting radial location of the wall differs from that of the radiative transfer simulation, we rerun the hydrodynamical simulation with the planet at a different radius, to make sure that the wall location between both simulations is self-consistent. The wall shapes and location before and after iteration are shown in Table \ref{tab:Tprof}.

For walls that were already quite round \textit{before} iterating on the scale height profile ($M>30 M_{\rm Jup}$ and $\alpha > 5\e{-3}$), using the calculated scale height from the radiative transfer model does not affect the wall shape significantly. For lower planet masses and disk viscosities, corresponding to steep walls before iteration, the wall shape does change. Iterating on the pressure profile makes these walls rounder by $\Delta w=0.04...0.07$. An example of this is shown in figure \ref{fig:Tprof}, for a 20 Jupiter mass planet in a moderately viscous disk ($\alpha = 2\e{-3}$). By looking at figure \ref{fig:contour}, this moves down the lower limit on the possible range of planet masses and disk viscosities from 30 to 20 Jupiter masses and from $\alpha = 5\e{-3}$ to $2\e{-3}$, respectively. 

We show that iterating on the scale height profile in hydrodynamical simulations does affect the resulting wall shapes. However, we have not taken the feedback of the disk structure on the pressure profile \textit{during} gap opening into account. To do this, one would need to recalculate the scale height profile during the hydrodynamical calculation, such that the wall shape and scale height profile are always self-consistent. However, such an approach is clearly beyond the scope of this paper. We leave it for future work, but we do note that the temperature structure in the wall may contribute to its roundness, and could play a role in other processes relevant to transitional disks such as the accretion flow across the gap. Ideally, one would also take into account the 3D structure of the wall, though this might take a considerable computational effort due to the long integration times necessary to reach a steady state.

\begin{figure}
  \includegraphics[width=\linewidth]{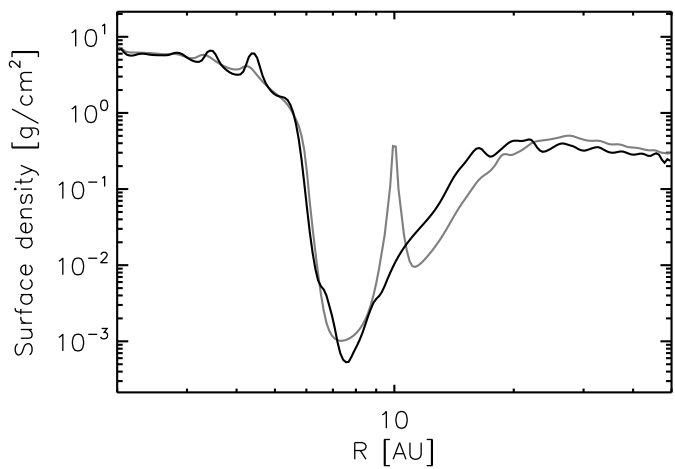}
  \caption[]{Comparison of the radial surface density profile (black line) with the azimuthally averaged one (grey line), for our best-fit model of figure \ref{fig:shape}.
    \label{fig:cut}}
\end{figure}

\subsection{Gap eccentricity}\label{sec:ellip}
Throughout this work, we have assumed azimuthal symmetry for the disk. However, towards the highest companion masses in our simulations, the gap starts to become eccentric (see also \citealt{2012ApJ...754L..31C}). In addition, the chromatic phases measured with MIDI also show indications for an asymmetric outer disk \citep{Panic:2012un}. Taking an azimuthal average over an eccentric gap can lead to a rounder surface density profile, however, we will show that this is not the effect we observe or measure in our simulations.

From a modelling point of view, making a gap eccentric contributes to the roundness measured from an azimuthally averaged surface density profile under certain conditions. For example, an elliptical gap ($e=0.4$) with a vertical wall has an azimuthally averaged surface density profile that appears slightly rounded-off (we fit $w=0.05$). This is much less round than the lowest roundness we measure in our simulations, which is $w=0.17$ (for a circular gap). To investigate if the eccentricity may contribute to the inferred roundness we compare a radial cuts of the surface density profile to azimuthally averaged ones. An example is shown in figure \ref{fig:cut} for our best-fit model, which has an eccentricity of $e \approx 0.6$. The radial cut is significantly more noisy -- motivating the use of azimuthally averaged profiles -- but the shape of the wall does not change significantly, though it is radially displaced by a few AU.

From an observational perspective, the eccentricity of the gap does not change the roundness of the disk wall as inferred from the visibilities. Each visibility is measured along a particular baseline orientation, and is therefore sensitive to the radial gradient of the surface density along that baseline, without any azimuthal averaging. The radial displacement of the disk wall due to eccentricity is measurable, but unfortunately this effect is degenerate with disk inclination and position angle, increasing the uncertainties on these parameters.

It may be possible to constrain the eccentricity from the the chromatic phases, hence providing an extra diagnostic on the companion mass, though this will require in-depth modeling of a more extended data set.

\section{Conclusion}\label{sec:conclusion}
We have studied the gap shape in the disk of \object{HD 100546}. By comparing 2D radiative transfer models to the mid-infrared interferometric data presented in \cite{Panic:2012un}, we find that:
\begin{itemize}
\item The disk wall at the far end of the gap is not vertical, but rounded off over a significant radial range. This shape can be explained by a gradual increase in the surface density over a range of $\sim$10 to $\sim$25 AU, creating a broad peak in the surface brightness profile around 12 AU that was also seen by \cite{Panic:2012un}.
\item The inner dust disk is extremely small  in small grains. We confirm the upper limit on its size of 0.7 AU found by \cite{Panic:2012un}, and use our 2D radiative transfer model to constrain it even further, with no measurable contribution outside of 0.3 AU.
\item The roundness or spatial extent of a disk wall can be inferred from \textit{spectrally} resolved visibilities beyond the first null.
\end{itemize}
By comparing these results to hydrodynamical simulations of planet-disk interactions, we find that:
\begin{itemize}
\item The shape of the surface density profile in the disk wall of HD 100546 can be explained by a massive planet ($\gtrsim$30...80 $M_{\rm Jup}$) in a viscous ($\alpha \gtrsim 5\cdot 10^{-3}$) disk between 8 and 10 AU.
\item The roundness of a disk wall in hydrodynamical simulations depends on the temperature structure \textit{in} the disk wall: iterating on the scale height profile using radiative transfer changes this roundness.
\item The disk viscosity is a crucial parameter in estimating planet masses from a derived surface density profile, acting against gap opening by the planet. For a given depth, the gap-opening mass of a planet increases with the square root of the disks viscosity.
\item The effect of a single planet is not enough to explain the full width of the disk gap in HD 100546. Either an additional clearing mechanism, or a multi-planet system is required to explain its extent.
\item The object shaping the disk wall is most likely a brown dwarf, suggesting HD100546 might be a binary system rather than a transitional disk object.
\end{itemize}

\begin{acknowledgements}
  This research project is financially supported by a joint grant from the
  Netherlands Research School for Astronomy (NOVA) and the Netherlands
  Institute for Space Research (SRON). 
  The work of O. P. was supported by the European Union through ERC grant number 279973.
\end{acknowledgements}
% 
% The bibliography
% 
\bibliographystyle{aa} % style aa.bst
%\bibliography{references} % using bibtool, see makepaper.sh line 2
\bibliography{../../Dropbox/papers/all,../../Dropbox/papers/books}
% 
% The appendix
% 
%\begin{appendix}
%\end{appendix}

\end{document}